\documentclass[%
 aip,
 amsmath,amssymb,
 reprint,%
]{revtex4-2}

\usepackage{graphicx}
\usepackage{dcolumn}
\usepackage{bm}
\usepackage[utf8]{inputenc}
\usepackage[T1]{fontenc}
\usepackage{mathptmx}
\usepackage{subfigure}
\usepackage{ulem}
\usepackage{xcolor}

\begin{document}

\preprint{APS/123-QED}

This is the accepted version of J. M.
De Ponti, E. Riva, F. Braghin, and R. Ardito, Elastic three-dimensional metaframe for selective wave filtering and polarization control, Appl. Phys. Lett. 119, 211903 (2021).
The final publication is available at:
\textcolor{blue}{https://doi.org/10.1063/5.0065553}

\title{Elastic three-dimensional metaframe for selective wave filtering and polarization control}

\author{J. M. De Ponti}
\email{jacopomaria.deponti@polimi.it}
\affiliation{Dept. of Civil and Environmental Engineering, Politecnico di Milano, Piazza Leonardo da Vinci 32 20133 Milano Italy}

\author{E. Riva}
\affiliation{Dept. of Mechanical Engineering, Politecnico di Milano, Via Giuseppe La Masa 1 20156 Milano Italy}

\author{F. Braghin}
\affiliation{Dept. of Mechanical Engineering, Politecnico di Milano, Via Giuseppe La Masa 1 20156 Milano Italy}

\author{R. Ardito}
\affiliation{Dept. of Civil and Environmental Engineering, Politecnico di Milano, Piazza Leonardo da Vinci 32 20133 Milano Italy}

\begin{abstract}
We experimentally achieve selective wave filtering and polarization control in a three-dimensional elastic frame embedding local resonators. By connecting multi-resonating elements to a frame structure, a complete low-frequency, subwavelength bandgap with strong  selective filtering properties is obtained. Theory and experiments demonstrate the metaframe capability to selectively stop transverse waves while allowing longitudinal wave propagation, as in 'fluid-like' elasticity. This peculiar behaviour, together with the complete bandgap structure, may open opportunities in the context of wave control, envisaging concurrent applications for three-dimensional filters and elastic wave polarizers.    
\end{abstract}

\maketitle

A dominant line of research in the wave physics focuses on the investigation of materials and structures to control the propagation of waves. Periodic media have been demonstrated particularly influential to manipulate waves exploiting Bragg scattering bandgaps, creating  the fields of photonic crystals in optics \cite{Joannopoulos2008}  and their acoustic counterparts in phononic structures \cite{Kushwaha1993,Laude2015,Hussein2014}. In the context of elastic waves, notable effort has been devoted to the definition of low frequency and subwavelength devices to comply with ambient and most widespread vibration spectra. To push the operational regime towards lower frequencies, the exploitation of local resonance has received considerable attention \cite{Liu2000,Ma2016,Miroshnichenko2010,Lemoult2011}, leading to the definition of metamaterials \cite{Craster2013Book,Craster2017Book}. Potential applications of elastic metamaterials could be implemented at any lengthscale, for seismic wave protection, \cite{Brule2014,Achaoui2017,Finocchio2014,Colombi2016a,Miniaci2016,Colombi2016b}, vibration absorption \cite{Colombi2017,Matlack2016,Kaina2013,Moscatelli2019}, nondestructive evaluation \cite{Moleron2015} and several nuances for wave enhancement and manipulation \cite{RomeroGarcia2013, Chen2014,Liu2017} recently inspired by topological insulators\cite{Mousavi2015, Susstrunk2015, Chaplain2020}. Metamaterials are often employed in combination with smart materials to obtain multifunctional devices for enhanced sensing or energy harvesting applications \cite{Sugino2018,DePonti2020,DePonti2021}.
The effect of local resonance on the dispersion properties in continuous elastic substrates has been widely investigated in the past through analytical and approximated formulations in beams \cite{Baravelli2013,Sugino2016}, plates \cite{Williams2015, Colquitt2017,Yoritomo2016} and halfspaces \cite{Colquitt2017,Comi2020}, amongst others.
Motivated by the idea to control elastic waves in a complete three-dimensional setting, we propose a metaframe able to provide selective wave filtering and polarization control. Moreover, we show that simple lumped models can be suitably employed to predict the behaviour of such structures, providing a rapid analytical design tool. We show that the attenuation performance of locally resonant structures strongly depends on the relative stiffness ratio (frame vs. resonator), which should be considered an additional parameter to the conventional total mass ratio \cite{Sugino2016}.
Even if a complete bandgap is obtained, the imaginary dispersion curves are remarkably different for longitudinal and transverse waves; this behavior is linked to the stiffness ratio between the frame and the resonator, which affects the attenuation properties of the structure.
More specifically, we demonstrate selective attenuation of transverse waves, while allowing longitudinal wave propagation, as in 'fluid-like' elasticity \cite{Ma2016, Lai2011}. The possibility to create polarization bandgaps by a selective suppression of each vibration mode has been demonstrated in stiff-in-soft structures with a single resonator per cell, obtaining different bandgaps for flexural, longitudinal or torsional modes \cite{Ma2016}, or leveraging multiple resonances \cite{Lai2011}. Modes can be also isolated on the basis of their polarization by creating distinct topological interfaces \cite{Miniaci2019} supporting heterogeneous helical-valley edge waves. Differently from these approaches, we show a three-dimensional metaframe with the coexistence of attenuation and trasmission of transverse and longitudinal waves, both inside a common bandgap.

\begin{figure}[h!]
\centering
    \includegraphics[width=0.5\textwidth]{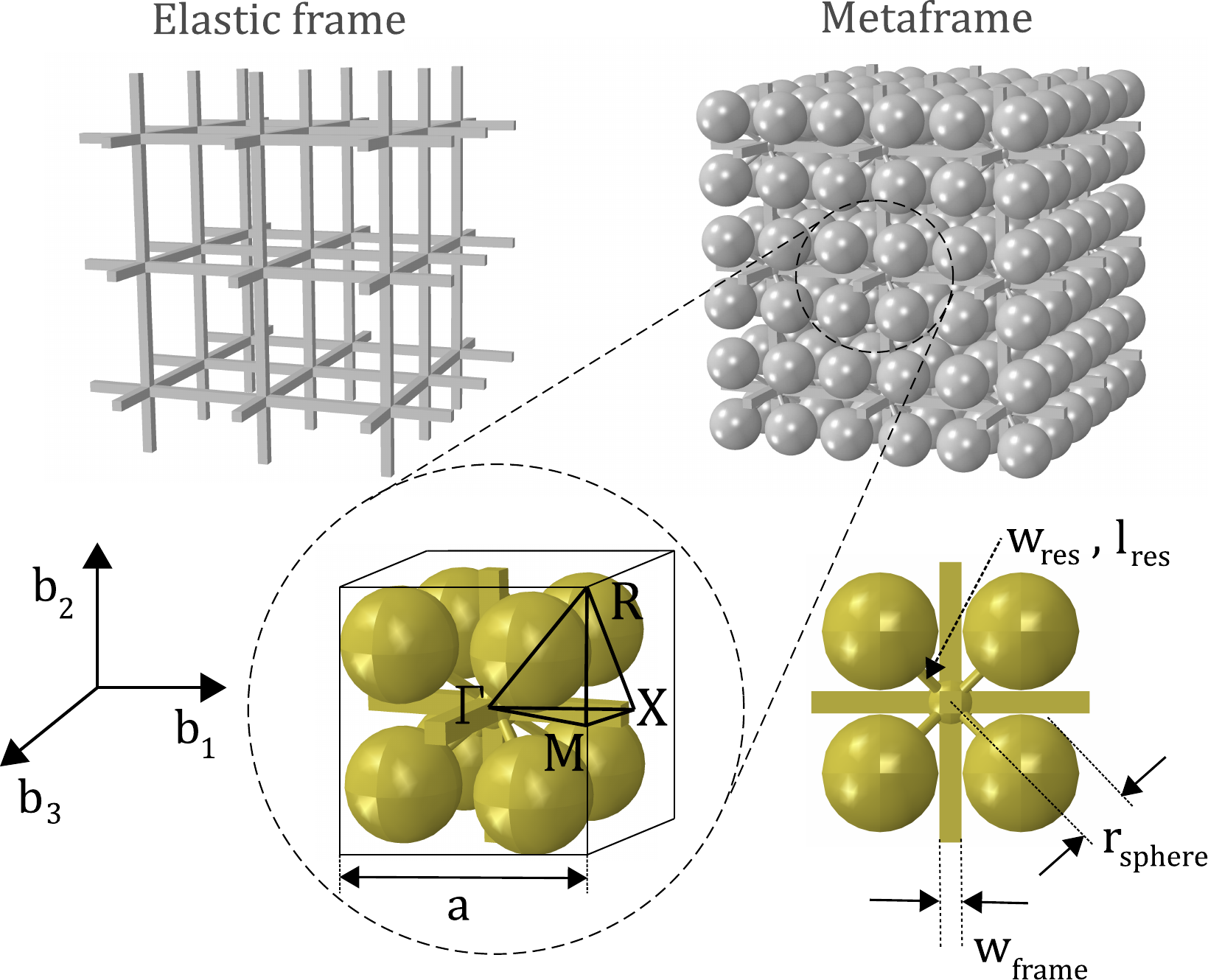}
   \caption{Elastic three-dimensional metaframe. The structure is obtained by connecting a set of resonators to the nodes of an elastic frame. In the bottom part, a detailed view of the unit cell made of a frame with 8 resonators attached is shown, together with the main geometrical dimensions and the high symmetry points used to compute the dispersion curves.}
	\label{fig:M1}
\end{figure}

\begin{figure*}[t!]
\centering
    \includegraphics[width=1\textwidth]{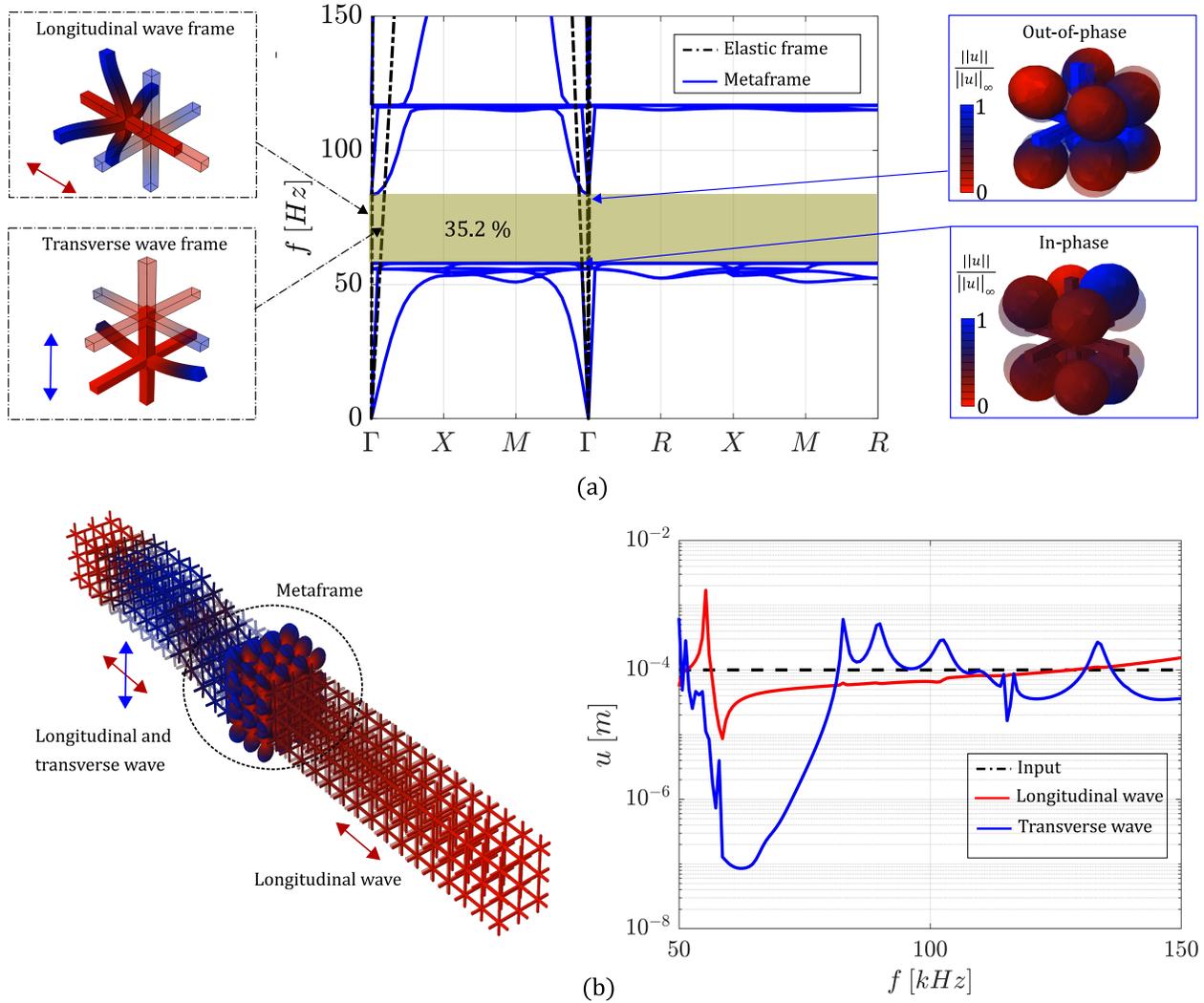}
   \caption{(a) Numerical dispersion curves of the metaframe obtained through FEM, superimposed to the dispersion of a plain elastic frame without the resonators. A  low frequency bandgap with gap-mid gap ratio of $35.2\%$ is observed in the range $58.04 - 83.28$ $Hz$ for both longitudinal and transverse waves.  Insets show the vibration modes of the plain frame and the metaframe at the $\Gamma$ point of the bandgap edges. The bandgap opens with an in-phase response of the resonators, while the closure is related to out-of-phase oscillations. (b) Wave propagation analyses on an infinitely long elastic frame embedding the metaframe. Selective wave filtering is observed: for the same input, longitudinal waves propagate, while transverse waves are attenuated.
   }
	\label{fig:M2}
\end{figure*}

Although this mechanism may resemble an elastic three-dimensional analogue of optical waveguide polarizers in photonics \cite{Sinha2006, Zimmermann2004}, the concept of wavelength selective filtering is caused here by a complete bandgap with an imaginary component strongly dependent on the wave polarization. This can be regarded as an additional feature coming from the higher complexity of wave polarization in elasticity, which allows creating complete bandgaps with strong selective filtering properties.
This mechanism can be employed for the creation of three-dimensional single phase selective filters, or elastic wave polarizers. \\ 

On the basis of previous research results \cite{DAlessandro2017}, we consider the metaframe depicted in Fig. \ref{fig:M1}, obtained connecting local resonators to an elastic frame. The finite structure is defined by periodically repeating a unit cell containing 8 resonators made of a thin elastic beam equipped with a sphere on the tip, and connected to a frame node. Such arrangement allows to increase the resonating mass, while keeping the stiffness of the resonators low.

The unit cell, of size $a=50$ mm, includes a frame with beams of square cross section of dimension $w_{frame}=0.08a$, while the resonators are characterized by a cylindrical elastic connection of diameter $w_{res}=0.04a$, length $l_{res}=0.14a$ and a sphere of radius $r_{sphere}=0.21a$. 
The entire structure is made of Nylon $PA12$, with Young's modulus $E_{PA12}=1586$ MPa, Poisson's ratio $\nu=0.4$ and density $\rho=1000$ kg/m$^3$.\\
The numerical dispersion curves of the metaframe, obtained through Abaqus\textsuperscript{\textregistered}, are displayed in Fig. \ref{fig:M2} (a), along with the graphical representation of a number of Bloch modes. A complete low- frequency bandgap in the range $58.04 - 83.28$ Hz, with gap-mid gap ratio \cite{DAlessandro2017} of $35.2\%$ is observed. In terms of normalized frequency, i.e. in units of the resonance frequency, we obtain a bandgap in the range $0.97 - 1.39$.  
In correspondence of the bandgap edges, the Bloch modes exhibit a localized flexural motion of the resonators, with in-phase and out-of-phase response for the mode below and above, respectively. At this step, we also observe that the number of propagating wave modes that collapse in a bandgap can be dynamically affected in a different manner from the resonator motion, due to the associated equivalent stiffness, which is very different for longitudinal and transverse waves. Fig. \ref{fig:M2} (b) shows the wave propagation along an infinitely long elastic frame with the insertion of the metaframe composed of 9 unit cells. Spurious edge reflections are avoided by imposing absorbing boundary conditions at the end of the frame along the wave propagation direction. They are implemented in Abaqus\textsuperscript{\textregistered} using the ALID (Absorbing Layers using Increasing Damping) method, adopting a cubic law function for mass proportional Rayleigh damping \cite{Rajagopal2012}.
By inspecting the wave amplitude after the metaframe it can be noticed that, for a constant input excitation, longitudinal waves propagate, while transverse waves are attenuated. This phenomenon is observed independently on the metaframe dimension, as can be noticed considering larger samples (see Supplementary Material). While the influence of different resonator modes on the attenuation performances has already been investigated \cite{Williams2015, Colquitt2017}, comparing for e.g. flexural and longitudinal resonances, we focus our attention on a single mode of the resonator, changing the stiffness of the frame only. This is possible thanks to the three-dimensional symmetry of the cell, which guarantees a flexural polarization of the resonators in accord with both longitudinal and transverse waves.

\begin{figure}[h!]
\centering
    \includegraphics[width=0.5\textwidth]{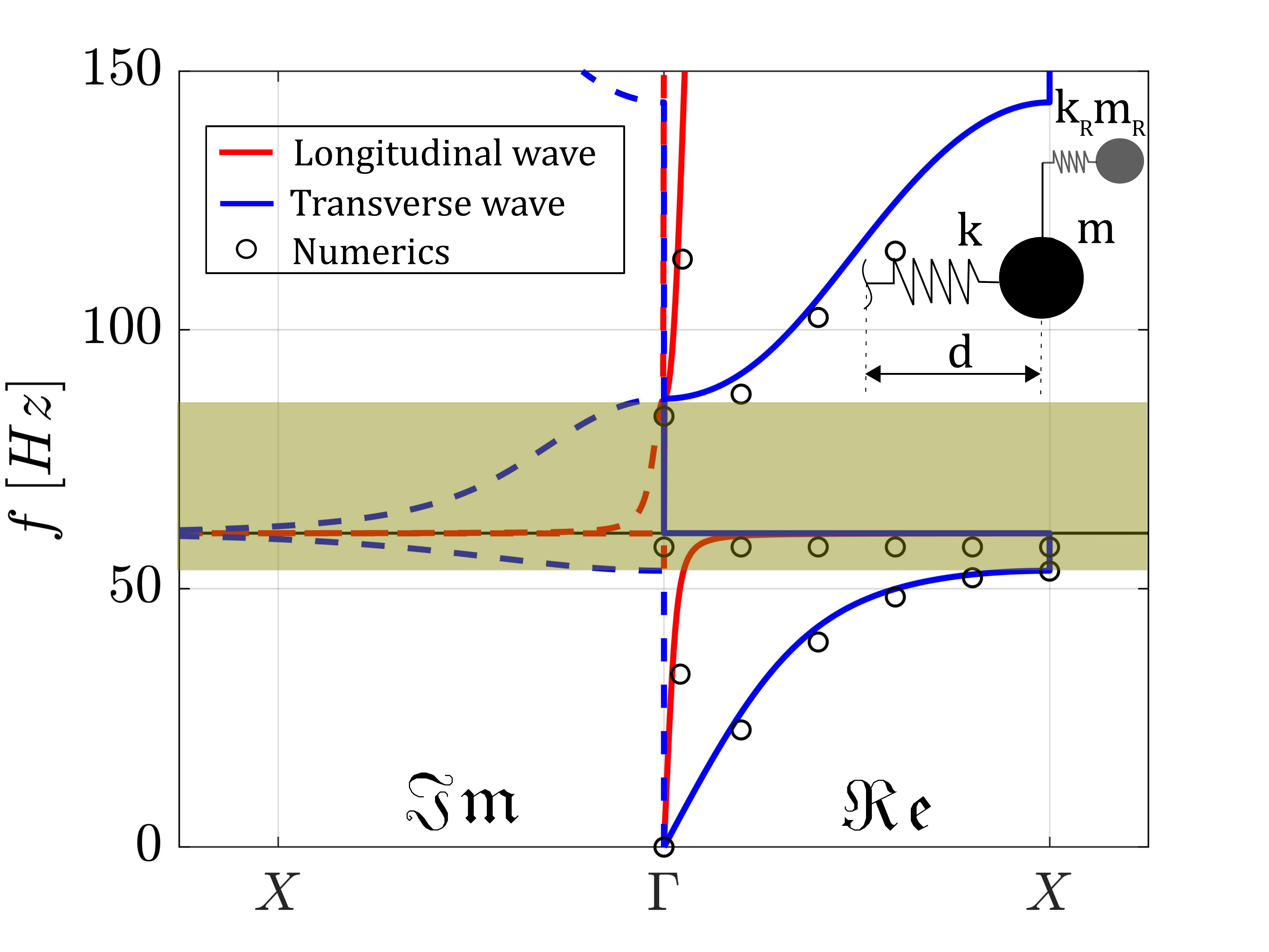}
   \caption{Analytical dispersion curve for the metaframe, superimposed to the numerical dispersion (scattered points) from FEM. Even if the bandgap exists for both longitudinal and transverse waves, the imaginary part of the dispersion curves is remarkably different, reporting almost negligible attenuation for longitudinal waves.}
	\label{fig:M3}
\end{figure}

To investigate on this aspect, we adopt a simplified lumped model to predict the bandgap characteristics, and the associated level of attenuation.
We consider a periodic spring-mass chain, made of unit cells of stiffness $k$ and mass $M$, with attached a resonator of stiffness $k_R$ and mass $m_R$, as shown in the inset of Fig. \ref{fig:M3}. The quantity $k$ represents the stiffness of the plain frame, while $m$ is the overall non-resonating mass. We adopt $2$ degrees of freedom to describe the translation of the main mass $M$ and of the resonator $m_R$. 

By defining $\Omega^2=k/M$, $\omega_R^2=k_R/m_R$, and introducing the non-dimensional quantity $\mu=\kappa d$, being $d$ the unit cell size and $\kappa$ the wavenumber, the dispersion relation is:
\begin{equation}
\cos(\mu) = 1-\frac{\omega^2}{2\Omega^2}\left[1+\frac{m_R}{M}\frac{\omega_R^2}{\omega_R^2-\omega^2}\right]\label{Eq.1}
\end{equation}
\noindent Evanescent waves are present if $\cos(\mu)\geq 1$ and $\cos(\mu) \leq -1$. By imposing these conditions we define, after some algebraic manipulations (see Supplementary Material), the limits of the bandgap as:
\begin{equation}
\scalebox{0.7}{$
\begin{split}
&\frac{4\Omega^2+\omega_R^2\left(1+\frac{m_R}{M}\right)-\sqrt{\left[4\Omega^2+\omega_R^2\left(1+\frac{m_R}{M}\right)\right]^2-16\omega_R^2\Omega^2}}{2}\leq\omega^2\leq\omega_R^2\left(1+\frac{m_R}{M}\right) \\ 
&\omega^2\geq \frac{4\Omega^2+\omega_R^2\left(1+\frac{m_R}{M}\right)+\sqrt{\left[4\Omega^2+\omega_R^2\left(1+\frac{m_R}{M}\right)\right]^2-16\omega_R^2\Omega^2}}{2}
\end{split}$}\label{Eq.2}
\end{equation}

The equations reveal that the bandgap does not open at the resonance frequency $\omega_R$, but at a frequency which depends also on the inertial and elastic properties of the main structure. 

The longitudinal (axial) and transverse (bending) stiffness of the frame can be easily computed by means of well-known techniques of structural mechanics (see Supplementary Material). Moreover, we model the resonator as a Timoshenko elastic beam with a lumped rigid mass, assuming as Lagrangian coordinates the translation of the tip of beam and the rotation of the mass. By doing so, we also neglect the mass of the beam ($0.46\%$ of the one of the sphere) and assume an infinite longitudinal (axial) stiffness.  By solving the eigenvalue problem for the $2$ degrees of freedom model of the resonator, we obtain that $82.4\%$ of the mass of the sphere participates to the motion, at the resonance frequency $f_{R}=60.7$  Hz. Having computed the mass and stiffness parameters, we get from Eq. \ref{Eq.1} the analytical dispersion curve reported in Fig.\ref{fig:M3}.
A good agreement between the lumped and the three-dimensional finite element model is observed. The simplified model provides accurate predictions in the frequency range of interest, with an error of approximately $4\%$ for the bandgap opening and closing frequencies. The plot shows clearly that the bandgap opening frequency for the transverse excitation is significantly smaller than the case of longitudinal wave. As a consequence, there is a frequency range in which transverse waves are evanescent and longitudinal waves are amplified; the relative width of such a range, with respect to the mid range frequency, is equal to $13 \%$. More importantly, the lumped model is used to easily evaluate the imaginary part of the dispersion relation, which quantifies the actual level of attenuation provided within the bandgap \cite{Laude2015}. It can be noticed that transverse waves are strongly attenuated, while the longitudinal ones are slightly affected by the bandgap, except for a narrow region centered at resonance frequency of the resonator. This feature allows the metaframe to selectively attenuate transverse waves and enable longitudinal wave propagation. That is, for a mixed longitudinal and transverse polarization of the input at the bandgap frequencies, one can make a filter for transverse wave components, while keeping unaltered the longitudinally polarized waves.
The different attenuation levels are attributed to the stronger mismatch between the axial stiffness of the frame and the flexural stiffness of the resonators. In terms of waves, a higher wavelength is obtained, thus increasing the subwavelength requirements of the system. \\ 

\begin{figure}[h!]
\centering
    \includegraphics[width=0.43\textwidth]{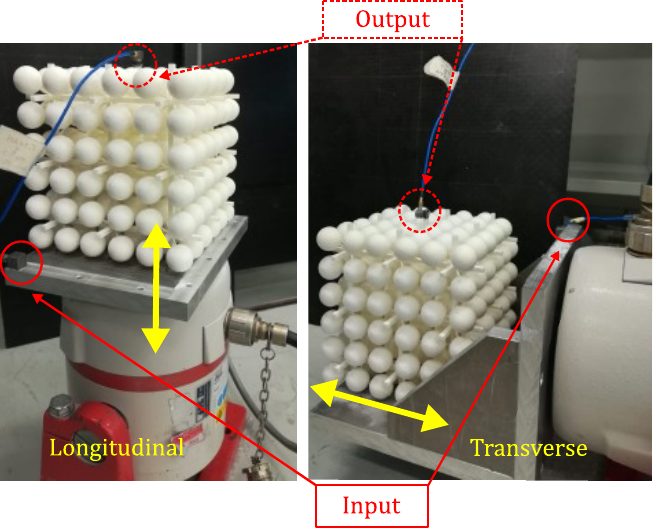}
   \caption{Experimental setup. The excitation is provided through an electrodynamic
shaker. The metaframe is glued to an aluminium thick plate which is rigidly connected to the shaker. Longitudinal excitation is provided by vertically placing the plate onto the shaker. For transverse excitation, the shaker is rotated of $90^{\circ}$ and the prototype is placed on a $L$-shaped fixture. Input and output accelerations are measured through accelerometers glued on the rigid plate and on the metaframe, respectively.}
	\label{fig:M4}
\end{figure}

The theoretical results are hereafter validated experimentally. Prototypes are built using Selective Laser Sintering (SLS), an additive manufacturing process that has no need of support material (a critical issue for the realization of suspended parts, like the resonators in this case) while guaranteeing good precision. The finite structure is composed of $9$ unit cells, resulting in a cube of dimension $150$ mm.  
The experimental setup is shown in Fig. \ref{fig:M4}.

The metaframe is glued to an aluminium thick plate which is connected to
a \textit{LDS v406} electrodynamic shaker to provide excitation. The aluminium plate is sufficiently thick to avoid relevant deformations for the input frequency, i.e. it is assumed as a perfectly rigid connection between the metaframe and the shaker.
Two configurations are tested, imposing a longitudinal and a transverse polarization of the wave respectively. Both cases consider the same direction of the wave propagation, $b_1$ in Fig.\ref{fig:M1}, but different polarization, i.e. $b_1$ for longitudinal waves and $b_3$ for transverse waves.
For the longitudinal wave, the plate is directly mounted onto the shaker which provides a vertical excitation. In contrast, the transverse wave is obtained by rotating the shaker of $90^{\circ}$ and connecting the plate to a $L$-shaped structure. The experimental acceleration spectra are measured using \textit{PCB Piezotronics 352C33} accelerometers with a sensitivity of $10$ mV/$g$ and resonant frequency higher than $70$ kHz. Since the accelerometers are able to measure the acceleration along one direction only, we rotated them of $90^{\circ}$ to measure transverse waves. The input is measured through an accelerometer glued to the thick plate, while the output is measured on the frame structure at the opposite side, as indicated by red circles and arrows in Fig. \ref{fig:M4}.
The excitation signal is provided through a \textit{KEYSIGHT
33500B} waveform generator, which synchronously starts with
the measurement system and consists in a linear swept-frequency cosine signal modulated by a Hann window. A frequency range between $50$ Hz to $150$ Hz is considered, filtering out the other frequency components. Data are then post-processed in order to estimate the transmission spectra, defined as $\tau\:$[dB] $=20$ log$_{10} (\ddot{w}_{OUT}/\ddot{w}_{IN})$.

\begin{figure}[h!]
\centering
    \includegraphics[width=0.5\textwidth]{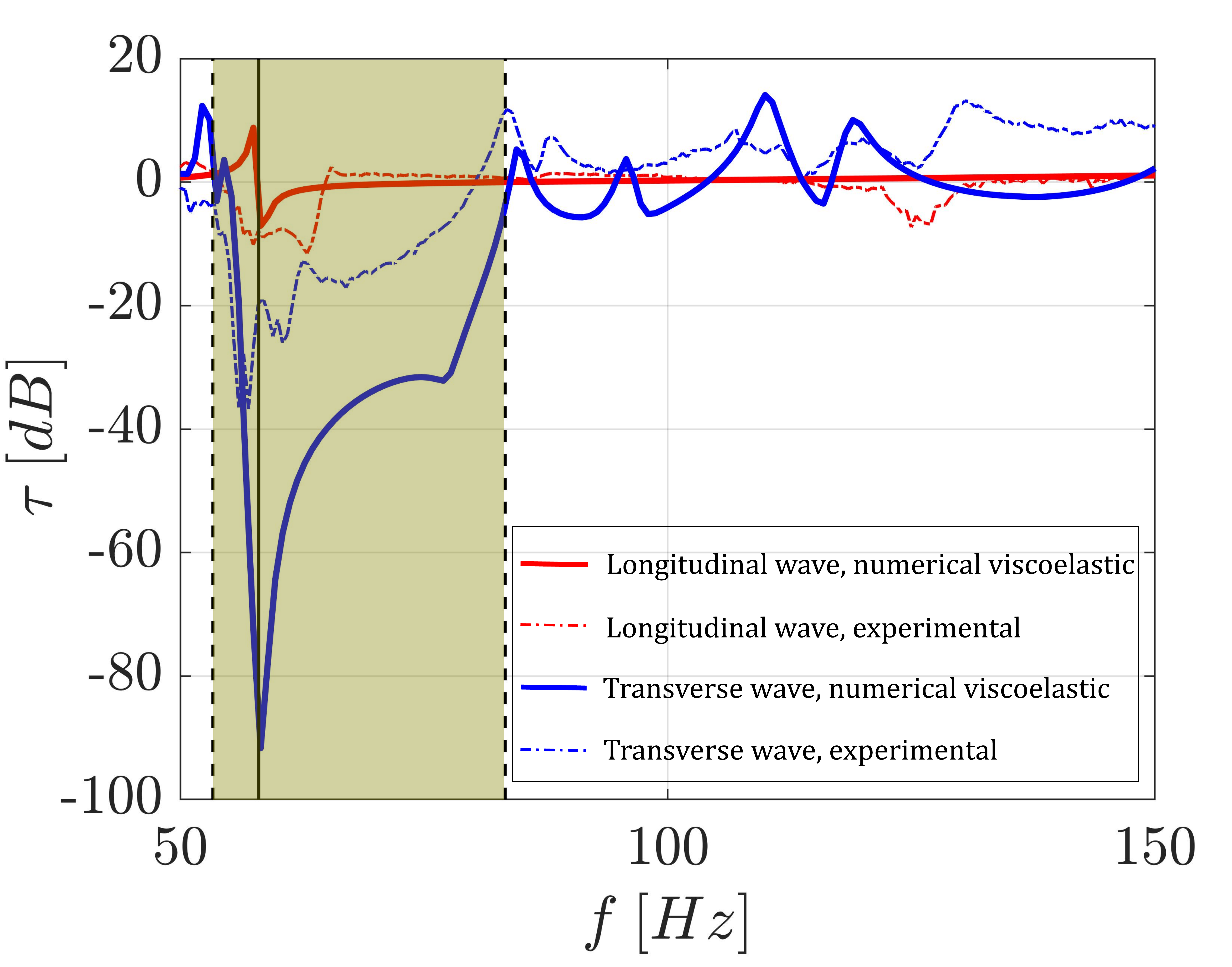}
   \caption{Numerical FEM viscoelastic (continuous lines) and experimental (dashed-dot lines) transmission spectra for longitudinal and transverse waves. Longitudinal waves (red lines) are slightly attenuated inside the expected bandgap region; maximum attenuation is detected at the resonance frequency of the resonators, marked by the continuous black vertical line. A stronger attenuation is measured for transverse waves (blue lines), in agreement with previous dispersion curves predictions.}
	\label{fig:M5}
\end{figure}

Fig. \ref{fig:M5} shows the numerical and experimental transmission diagrams for longitudinal and transverse waves. To investigate the results, we carry out FEM viscoelastic transmission analyses \cite{Krushynska2016,DAlessandro2020}, using the Standard Linear Solid – Maxwell type – model with $\tau_{Maxwell}=1.5e-4$ s and a Young’s modulus $E_{Maxwell}=315 \:$MPa. The experimental data are in satisfactory agreement with the computations, corroborating the different attenuation performance of the structure for longitudinal and transverse waves. The different level of attenuation, and the slight shift in the position of the transmission peaks is attributed to the relative simplicity of the viscoelastic model with respect to the actual material behaviour, as well as experimental misalignments  that prevent the possibility to excite longitudinal and transverse waves separately.

In conclusions, we have experimentally demonstrated potential advantages in using a three-dimensional metaframe for the creation of a complete, low-frequency, subwavelength bandgap. The strong mismatch between the longitudinal and transverse stiffness of the frame allows to achieve a different control of the waves depending on the polarization direction. Strong attenuation of transverse waves is achieved, while longitudinal wave propagation is mostly unaffected by the bandgap, except from a sharp region close to the resonant frequency of the resonators. This behavior can be suitably employed for applications involving  selective wave filtering and polarization control.

\section*{Supplementary Material}
See  the \textbf{supplementary material} for a detailed description of the analytical and numerical models.

\section*{Acknowledgements}
The  support  of  the  H2020  FET-proactive  project  MetaVEH  under  grant  agreement  No.  952039  is acknowledged. We also gratefully acknowledge the Italian Ministry of Education, University and Research for the support provided through the Project “Department of Excellence LIS4.0—Lightweight and Smart Structures for Industry 4.0.

\section*{Data Availability}
The data that support the findings of this study are available from the corresponding author upon reasonable request.

\end{document}


\preprint{APS/123-QED}

\title{Elastic three-dimensional metaframe for directional wave filtering and polarization control: Supplementary Material}

\author{J. M. De Ponti}
\email{jacopomaria.deponti@polimi.it}
 \affiliation{Dept. of Civil and Environmental Engineering, Politecnico di Milano, Piazza Leonardo da Vinci 32 20133 Milano Italy}
 
\author{E. Riva}
\affiliation{Dept. of Mechanical Engineering, Politecnico di Milano, Via Giuseppe La Masa 1 20156 Milano Italy}

\author{F. Braghin}
\affiliation{Dept. of Mechanical Engineering, Politecnico di Milano, Via Giuseppe La Masa 1 20156 Milano Italy}

\author{R. Ardito}
 \affiliation{Dept. of Civil and Environmental Engineering, Politecnico di Milano, Piazza Leonardo da Vinci 32 20133 Milano Italy}

\maketitle

\beginsupplement

To aid  insight  into the underlying physics of the elastic metaframe, described in the main text, we give a detailed description of the analytical and numerical models.

\section{SUPPLEMENTARY NOTE 1 - EQUIVALENT PROPERTIES OF THE RESONATOR}

The aim of this section is to show how the equivalent stiffness and participating mass of the resonators are evaluated. Such parameters are employed in the main text to compute the dispersion relation of the system through a simplified model.\\
We simplify the resonators by using a $2$ degrees of freedom lumped model. The elastic connection is modeled using Timoshenko beam theory, while the tip sphere is considered as a rigid mass.

\begin{figure}[h!]
\centering
    \includegraphics[width=0.4\textwidth]{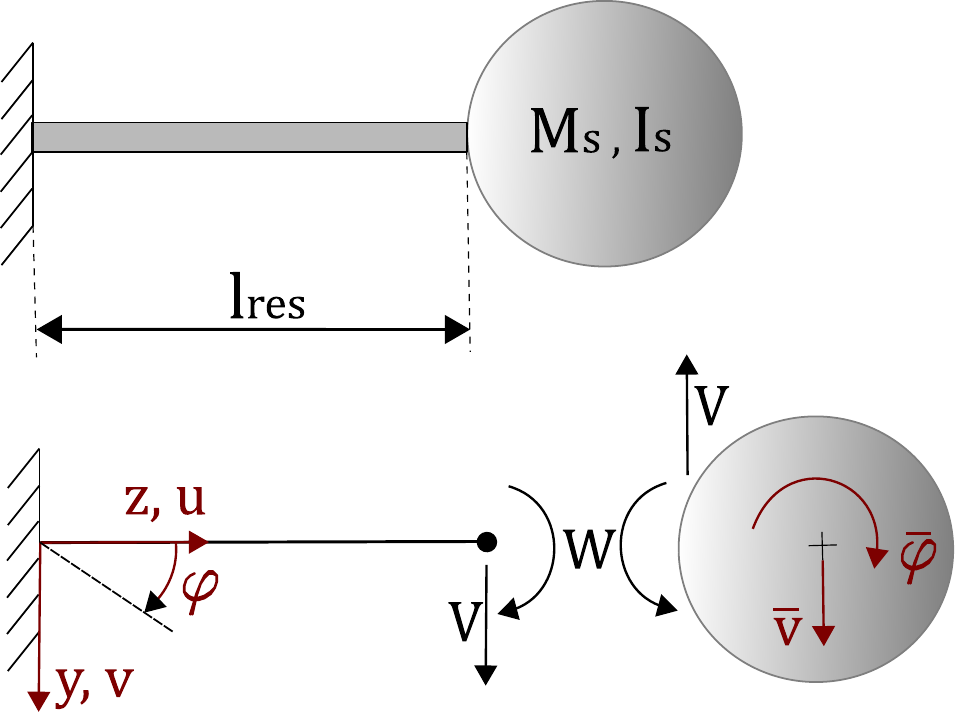}
   \caption{Schematic of the resonators attached to the elastic frame.}
	\label{fig:S1}
\end{figure}

Assuming a negligible mass of the beam with respect to the sphere, the governing equilibrium equations for the beam are:

\begin{equation}
    \begin{cases}
    \dfrac{d^2v}{dz^2}-\dfrac{d\varphi}{dz}=0\\
    EI\dfrac{d^2\varphi}{dz^2}+GA^{*}\left(\dfrac{dv}{dz}-\varphi\right)=0
     \end{cases}
\end{equation}\label{S1}
where $\bar{v}$ and $\bar{\varphi}$ are the translation and rotation of the mass respectively, with kinematic and static boundary conditions:

\begin{equation}
    \begin{cases}
    v\big|_{z=0}=0\\
    \varphi\big|_{z=0}=0\\
    \dfrac{d\varphi}{dz}\big|_{z=l}=-\dfrac{W}{EI}\\
    \left(\dfrac{dv}{dz}-\varphi\right)\bigg|_{z=l}=\dfrac{V}{GA^{*}}
     \end{cases}
\end{equation}\label{S2}

where $E$ is the material Young modulus, $I$ the second moment of area, $A^{*}$ the shear area and $G$ the shear modulus. By enforcing the boundary conditions we get:
\begin{equation}
\begin{split}
& v=-\frac{V}{6EI}z^3+\left(\frac{Vl}{2EI}-\frac{W}{2EI}\right)z^2 + \frac{V}{GA^{*}}z\\
& \varphi = -\frac{V}{2EI}z^2+\left(\frac{Vl}{EI}-\frac{W}{EI}\right)z
\end{split}\label{S3}
\end{equation}

The equilibrium equations for the rigid sphere are:

\begin{equation}
    \begin{cases}
       M_s\dfrac{d^2\bar{v}}{dt^2}+V = 0\\
       I_s\dfrac{d^2\bar{\varphi}}{dt^2}+M_sR\dfrac{d^2\bar{v}}{dt^2} + W = 0
    \end{cases}\label{S4}
\end{equation}

\noindent being $\bar{\varphi}=\varphi(l)$ and $\bar{v}=v(l)+\varphi(l)R$, with $R$ the sphere radius.
By expressing the unknown static quantities $V$ and $W$ in terms of $\varphi(l)$ and $v(l)$ from Eq. \ref{S3}, we write Eq. \ref{S4} as:
\begin{equation}
    \begin{cases}
       M_s\dfrac{d^2v(l)}{dt^2}+M_sR\dfrac{d^2\varphi(l)}{dt^2}+K_v v(l)-\dfrac{K_v l}{2}\varphi(l)=0\\
       M_sR\dfrac{d^2v(l)}{dt^2}+\left(I_s+M_sR^2 \right)\dfrac{d^2\varphi(l)}{dt^2}-\dfrac{K_v l}{2}v(l)+K_{\varphi} \varphi(l)=0
    \end{cases}\label{S5}
\end{equation}
with $K_v =(12EI/l^3)/(1+12EI/GA^{*}l^2)$ and $K_{\varphi}=(3EI/l)/(1+12EI/GA^{*}l^2)+EI/l$. Imposing harmonic solutions of the form $v(l)=\tilde{V}e^{i\omega t}$ and $\varphi(l)=\tilde{\Phi}e^{i\omega t}$, the following eigenvalue problem is obtained:

\begin{equation}
\left\{
\begin{bmatrix}
M_s  & M_sR \\
M_sR & (I_s+M_sR^2)\\
\end{bmatrix}\omega^2-
\begin{bmatrix}
K_v             & -\dfrac{K_v l}{2}\\
-\dfrac{K_v l}{2} & K_{\varphi}\\
\end{bmatrix}
\right\}
\begin{bmatrix}
\tilde{V}\\
\tilde{\Phi}\\
\end{bmatrix}=0
\end{equation}
By solving the eigenvalue problem, we finally get the resonance frequencies of the resonators:
\begin{equation}
\scalebox{0.9}{$
    \begin{split}
        & \omega^2 = \frac{2M_sR\frac{K_v l}{2}+K_v (I_s+M_sR^2)+K_{\varphi}M_s}{2M_sI_s}\pm\\
        & \frac{\sqrt{
               \begin{Bmatrix}
                                    [2M_sR\frac{K_v l}{2}+K_v (I_s+M_sR^2)- K_{\varphi}M_s]^2 + \\ +4K_v K_{\varphi}M_s^2R^2+4\frac{K_v l}{2}M_s(\frac{K_v l}{2}I_s+2M_sRK_{\varphi})
               \end{Bmatrix}}
               }{2M_sI_s}
    \end{split}
    $}
\end{equation}
and the eigenvectors $\bar{V}$ and $\bar{\Phi}$ corresponding to the obtained eigenvalues.
By doing so, we finally get the modal stiffness and modal mass:
\begin{equation}
\begin{cases}
k_{mod}= K_v\bar{V}^2-2\frac{k_vl}{2}\bar{V}\bar{\Phi}+K_{\varphi}\bar{\Phi}^2\\
m_{mod}=M_s\bar{V}^2+2M_sR\bar{V}\bar{\Phi}+(I_s+M_sR^2)\bar{\Phi}^2
\end{cases}
\end{equation}
and the modal participation factor:
\begin{equation}
\Gamma=\frac{M_s\bar{V}+M_sR\bar{\Phi}}{m_{mod}}
\end{equation}
Finally, the resonating mass is defined as:
\begin{equation}
m_{res} = m_{mod}\Gamma^2\\
\end{equation}

\section{SUPPLEMENTARY NOTE 2 - ANALYTICAL BANDGAP PREDICTION}
\begin{figure}[h!]
\centering
    \includegraphics[width=0.3\textwidth]{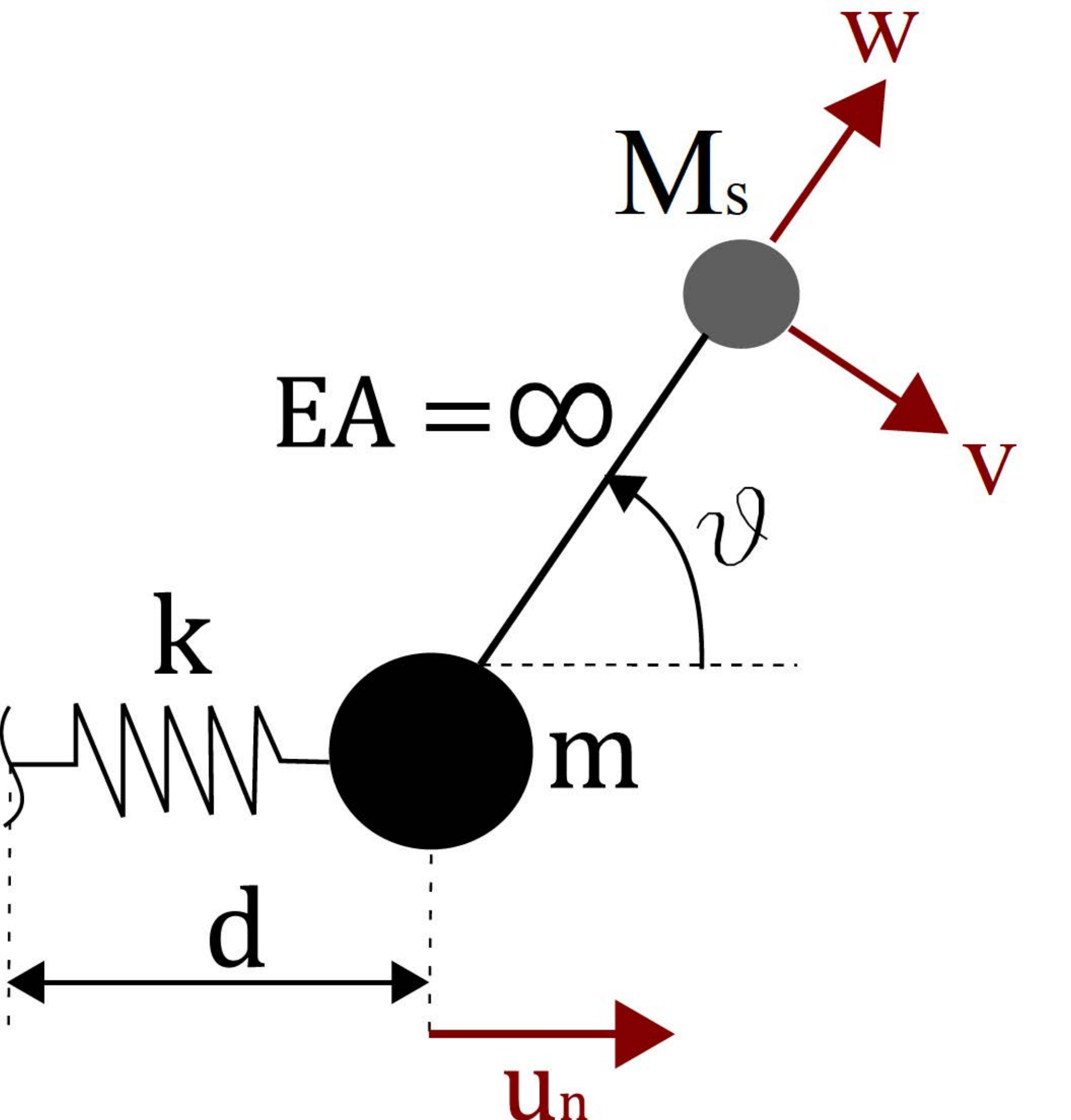}
   \caption{Schematic of the lumped spring-mass chain model used to predict the local resonance bandgap.}
	\label{fig:S2}
\end{figure}

We model the metaframe using a simple spring-mass chain, as depicted in Fig. \ref{fig:S2}.
The stiffness $k$ of the main structure is the axial or flexural stiffness of the frame depending on the polarization of the wave, while for the resonators we consider only the flexural stiffness, since the axial stiffness is assumed infinite. $m$ represent the lumped mass of the frame.

The equation governing the movement of the primary spring-mass chain is given by : 
\begin{multline}\label{S11}
m\ddot{u}_n = k(u_{n+1}-u_n)+k(u_{n-1}-u_n) +\\ +\sum_j V_j  \sin(\theta_j)+\sum_j H_j \cos(\theta_j)
\end{multline}

\noindent where $V_j$ and $H_j$ are the transverse and axial base reaction, respectively, for the j-th resonator attached to each node of the spring-mass chain.

Each resonator is subject to the base motion, projected in the transverse and axial directions:
\begin{equation}
\begin{split}
& v_{bj}=u_n\sin(\theta_j) \\
& u_{bj}=u_n\cos(\theta_j)
\end{split}
\end{equation}\label{basemotion}

By assuming infinite axial stiffness, one gets:
\begin{equation} \label{S13}
H_j = -M_s \ddot{u_n} \cos(\theta_j) 
\end{equation}

The transverse reaction is obtained on the basis of the analysis of the resonator. Eq. \ref{S5} is written in matrix form, with the addition of the base motion along the transverse direction only $\mathbf{U}_b = \mathbf{T} u_n \sin{\theta_j}$; $\mathbf{T}=[1 \quad 0]^T$ is the projecting vector.
\begin{equation}
\mathbf{M} \ddot{\mathbf{U}} +  \mathbf{K} \mathbf{U} = - \mathbf{M} \ddot{\mathbf{U}}_b 
\end{equation}

\noindent The reaction force is given by:
\begin{equation}
V_j=-\mathbf{T}^T \mathbf{M} (\ddot{\mathbf{U}} + \ddot{\mathbf{U}}_b) 
\end{equation}

\noindent The modal projection is adopted $\mathbf{U}=\boldsymbol{\Psi} q$, using the first dominant mode of the resonator and the relevant eigenvector $\boldsymbol{\Psi}$ as computed in the previous Section. As a consequence, one gets:
\begin{equation}\label{S16}
V_j=-m_{mod} \Gamma \ddot{q} - M_s \ddot{u_n} \sin(\theta_j) 
\end{equation} 

The solution is sought in the form $u_n=\tilde{u} e^{i(\kappa x_n - \omega t)}$, so that the modal response of the resonator reads:
\begin{equation}
q=\frac{\omega^2}{\omega^2_R - \omega^2} \Gamma \sin{\theta_j} \tilde{u} e^{i(\kappa x_n - \omega t)} 
\end{equation} 

After some algebra, the equation of motion \ref{S11} becomes:
\begin{multline}
-m \omega^2 \tilde{u} = k(e^{i \kappa d}+ e^{-i \kappa d}-2)+ \\ +\omega^2 \left( m_{res}\frac{\omega^2}{\omega^2_R - \omega^2} + M_s \right) \sum_j \sin^2(\theta_i) \tilde{u} + \omega^2 M_s \sum_j \cos^2(\theta_j) \tilde{u}
\end{multline}

\noindent The second addend in the r.h.s can be modified by defining the non-resonating mass as $m_{nres} = M_s - m_{res}$. Then, the dispersion relation is written as follows:
\begin{multline}
\cos(\kappa d) = 1 - \frac{\omega^2}{2k}\left[m+m_{nres} \sum_j \sin^2(\theta_j) + M_s \sum_j \cos^2(\theta_j)\right] +\\ -\frac{\omega^2}{2k} \frac{\omega_R^2}{\omega^2_R - \omega^2} m_{res} \sum_j \sin^2(\theta_j)
\end{multline}

The dispersive behavior of the spring-mass chain is dominated by the effective values of non-resonating mass and resonating mass. The former value is given by the mass of the frame added to the non-resonating mass for flexural motion of the resonator and to the axial contribution:
\begin{equation}
M=m+m_{nres} \sum_j \sin^2(\theta_j) + M_s \sum_j \cos^2(\theta_j)
\end{equation} 

\noindent The effective resonating mass is:
\begin{equation}
m_R=m_{res} \sum_j \sin^2(\theta_j)
\end{equation} 

\noindent For the specific geometry considered in this paper, one finds that $\sum_j \sin^2(\theta_j) = 5.33$ and  $\sum_j \cos^2(\theta_j) = 2.67$.

By defining $\Omega^2=k/M$ and introducing the non-dimensional quantities $\bar{\omega}^2=\omega^2/\Omega^2$, $\bar{\omega}_R^2=\omega_R^2/\Omega^2$, $\mu=\kappa d$, the dispersion relation is:
\begin{equation}
\cos(\mu) = 1-\frac{\bar{\omega}^2}{2}\left[1+\frac{m_R}{M}\frac{\bar{\omega}_R^2}{\bar{\omega}_R^2-\bar{\omega}^2}\right]\label{Eq.1}
\end{equation}
\noindent Evanescent are supported if (i) $\cos(\mu)\geq 1$ and (ii) $\cos(\mu) \leq -1$.
Condition (i) gives:

\begin{itemize}
    \item[-] if $\bar{\omega}^2<\bar{\omega}_R^2$ $\Rightarrow$ $\nexists$ $\bar{\omega}$
    \item[-] if $\bar{\omega}^2>\bar{\omega}_R^2$ $\Rightarrow$  $\bar{\omega}_R^2<\bar{\omega}^2<\bar{\omega}_R^2\left[1+\frac{m_R}{M}\right]$
\end{itemize}

Condition (ii) gives:

\begin{itemize}
    \item[-] if $\bar{\omega}^2<\bar{\omega}_R^2$ $\Rightarrow$ $\bar{\omega}_1^2<\bar{\omega}^2<\bar{\omega}_R^2$
    \item[-] if $\bar{\omega}^2>\bar{\omega}_R^2$ $\Rightarrow$  $\bar{\omega}^2>\bar{\omega}_2^2>\bar{\omega}_R^2$
\end{itemize}
where:
\begin{equation}
\begin{split}
& \bar{\omega}_1^2=\frac{4+\bar{\omega}_R^2(1+\frac{m_R}{M})-\sqrt{[4+\bar{\omega}_R^2(1+\frac{m_R}{M})]^2-16\bar{\omega}_R^2}}{2} \\
& \bar{\omega}_2^2=\frac{4+\bar{\omega}_R^2(1+\frac{m_R}{m})+\sqrt{[4+\bar{\omega}_R^2(1+\frac{m_R}{M})]^2-16\bar{\omega}_R^2}}{2}
\end{split}
\end{equation}\

\noindent In conclusion, evanescent waves exist in the range $[\bar{\omega}_1, \bar{\omega}_R\sqrt{1+m_R/M}]$ and $[\bar{\omega_2}, +\infty)$.
 
We sum up in table \ref{Tab:S1} the main elastic and inertial properties of the resonator and the main structure used to obtain the analytical dispersion curves. The last row contains the effective non-resonating and resonating mass, computed on the basis of the aforementioned definition and of the geometry of the considered metaframe.

\begin{table}[h!]
\centering
\begin{tabular}{ c  c  c  c }
\hline
 & Frame & Resonators  &\\
\hline
 $k_{axial}\:[N/mm]$ & $507.520$ & $\infty$ \\
\hline
 $k_{flexural}\:[N/mm]$ & $3.248$ & $34.412$ \\
\hline
 $m_{no-res./res.} \:[g]$ & $20.471$ & $21.297$ \\
\hline
\end{tabular}
\caption{Elastic and inertial parameters adopted in the analytical lumped model.}
\label{Tab:S1}
\end{table}

\section{SUPPLEMENTARY NOTE 3 - ADDITIONAL FEM ANALYSES}

\begin{figure}[h!]
\centering
    \includegraphics[width=0.55\textwidth]{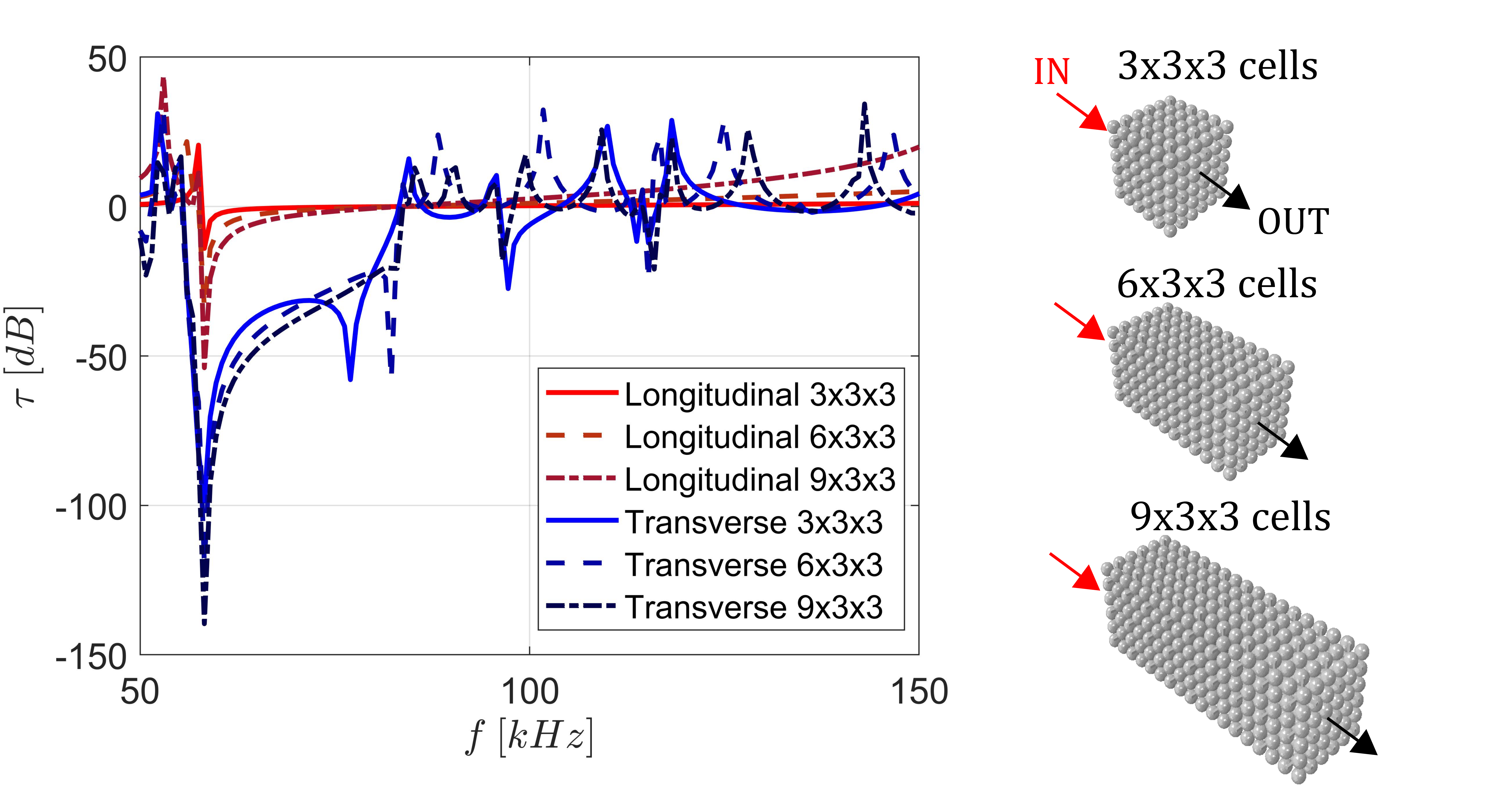}
   \caption{Transmission analyses for 3, 6 and 9 cells along the wave propagation direction. Even increasing the number of cells, a remarkably different behaviour for longitudinal and transverse waves is observed.}
	\label{fig:S3}
\end{figure}

To validate the size-independent behaviour of the proposed elastic metaframe, we perform FEM analyses on larger samples, with 6 and 9 cells along the propagation direction of the wave. Fig. \ref{fig:S3} shows the response of the metaframe for increasing number of cells along the wave propagation direction. It can be noticed that increasing the number of cells, transverse waves still are much more attenuated than longitudinal waves. Such behaviour totally complies with the imaginary part of the dispersion relation, which shows a different attenuation for transverse and longitudinal waves.

\section{SUPPLEMENTARY NOTE 4 - NUMERICAL FEM MODEL}
The numerical FEM model is carried out using the commercial software Abaqus\textsuperscript{\textregistered}. The metaframe is modeled as a deformable three-dimensional structure which is discretized through a free mesh made of 10-node quadratic tetraedra (C3D10). The dispersion curves are obtained through a modal analysis of the unit cell complemented with a Matlab code to apply Bloch-Floquet boundary conditions. Transmission spectra are obtained performing steady-state dynamics direct analyses, in which the viscoelastic model is introduced by means of Prony series relaxation functions, defined by the coefficients $g_i$ and $\tau_i$, that Abaqus\textsuperscript{\textregistered} relates to the storage $G_s(\omega)$ and loss moduli $G_l(\omega)$. \cite{Abaqus}